\documentclass[preprint, showpacs,amsmath,amssymb]{revtex4}

\usepackage{graphicx}
\usepackage{dcolumn}
\usepackage{bm}

\begin{document}


\title{Dynamics of intracellular Ca$^{2+}$ oscillations in the presence of multisite Ca$^{2+}$-binding proteins}

\author{Roberto~Chignola}
\affiliation{Dipartimento di Biotecnologie, Universit\`a di Verona, Strada Le Grazie 15 Ð CV1, I-37134 Verona, Italy}
\affiliation{Istituto Nazionale di Fisica Nucleare -- Sezione di Trieste, Via Valerio 2, I-34127 Trieste, Italy}
 
\author{Alessio Del Fabbro}%
\affiliation{Universit\`a di Trieste and Istituto Nazionale di Fisica Nucleare -- Sezione di Trieste, Via Valerio 2, I-34127 Trieste, Italy}%

\author{Edoardo~Milotti}
\affiliation{Universit\`a di Trieste and Istituto Nazionale di Fisica Nucleare -- Sezione di Trieste, Via Valerio 2, I-34127 Trieste, Italy}%
\email{milotti@ts.infn.it}
\homepage{http://www.ts.infn.it/physics/experiments/vbl/SiteVBL/index.html}

\date{\today}

\begin{abstract}
We study the dynamics of intracellular calcium oscillations in the presence of proteins that bind calcium on multiple sites and that are generally believed to act as passive calcium buffers in cells. We find that multisite calcium-binding proteins set a sharp threshold for calcium oscillations. Even with high concentrations of calcium-binding proteins, internal noise, which shows up spontaneously in cells in the process of calcium wave formation, can lead to self-oscillations. This produces oscillatory behaviors strikingly similar to those observed in real cells. In addition, for given intracellular concentrations of both calcium and  calcium-binding proteins the regularity of these oscillations changes and reaches a maximum as a function noise variance, and the overall system dynamics displays stochastic coherence.  We conclude that calcium-binding proteins may have an important and active role in cellular communication.
\end{abstract}

                             
\maketitle
Animal cells communicate through the exchange of molecules secreted in the extracellular environment and/or by cell-cell contacts through the chemical interactions of molecules expressed at the cell surface.  Cell communication is at the basis of animal physiology and pathology, and the information stored in the  environment must ultimately be decoded in the cell and propagate to the nucleus where target genes can either switch on or off in response to these stimuli.  Calcium (Ca$^{2+}$) is the most versatile second messenger in living cells and translates the information stored in the extracellular environment in time-dependent variations of Ca$^{2+}$ intracellular concentration. These may take the form of waves, bursts and oscillations that propagate in time and space in the cell and through adjacent cells \cite{Berridge98, Carafoli01, Falcke04}.

Ca$^{2+}$ oscillations occur in a large number of cell types such as excitable (e.g. neurons, cardiac cells) and non excitable (e.g. hepatocytes, endothelial cells) cells, either spontaneously or after stimulation by hormones, cytokines and neurotransmitters, and they drive important functions such as brain and cardiac activity, immune cell activation, hormone secretion and cell death \cite{Berridge98, Carafoli01}.

Two aspects of intracellular Ca$^{2+}$ oscillation have received little attention, with some notable exceptions \cite{Falcke03A, Falcke03B}:  Ca$^{2+}$ oscillations propagate in the cell in the presence of Ca$^{2+}$-binding proteins that act as buffers, taking up to 99\% of Ca$^{2+}$ in the cell; the mechanisms of Ca$^{2+}$ wave generation is intrinsically noisy. Here we address both aspects from a dynamical perspective.

Ca$^{2+}$ concentration within cells is controlled by reversible binding to specific classes of proteins that act as Ca$^{2+}$ sensors and decode the information carried by FM and AM modulation of the Ca$^{2+}$ oscillations \cite{sch}. Many intracellular proteins can bind Ca$^{2+}$ on multiple sites.  For example, the activity of key enzymes such as CaM-kinase II is modulated either by direct binding to Ca$^{2+}$ or indirectly by chemical interaction with Ca$^{2+}$-binding proteins \cite{Carafoli01}. Ca$^{2+}$ binding to proteins on multiple sites is not driven by enzymatic mechanisms and therefore its chemistry obeys to the law of mass action. In addition, the binding sites on Ca$^{2+}$-binding proteins appear to be all equivalent in presence of the EF hand motif which allows the specific and reversible binding of Ca$^{2+}$ ions \cite{Carafoli01}.\\
In this paper we investigate Ca$^{2+}$ dynamics in cells in the presence of Ca$^{2+}$-binding proteins and noise.
To this end, the vitamin D-dependent protein calbindin-D29K (CaL) has been chosen as an example of multisite Ca$^{2+}$ binding protein, representing a class of several Ca$^{2+}$ binding proteins that share many biochemical traits with CaL \cite{Carafoli01}.
CaL is a protein with a molecular weight of 30 kDa,  that is expressed in the cytosol of specific cells of the central nervous system such as hippocampal granule cells and cerebellar Purkinje cells \cite{Carafoli01}. CaL is a member of the EF hand superfamily  Ca$^{2+}$-binding proteins. To date, CaL has not yet been found to play a role in Ca$^{2+}$-dependent regulation of enzyme activity although it is expected to act as an intracellular Ca$^{2+}$ buffer. This seemingly minor role may nonetheless have a profound effect on cell physiology, and indeed overexpression of CaL in cultured hippocampal pyramidal neurons affects synaptic plasticity and suppressed post-tetanic potentiation \cite{Chard95}. CaL has four reversible binding sites for Ca$^{2+}$ and, importantly, the on-off rates for the binding reaction have been determined experimentally. Experiments have also shown the presence of binding sites with two different affinities and a ratio of sites with high and low affinity of  3 to 1 or 2 to 2 \cite{Nageri00}.

We describe the dynamics of Ca$^{2+}$ reversible binding to CaL with the following differential system, which is formally the same that we used in a previous study of multisite protein modification  (\cite{Chignola06,Milotti07}; see also \cite{rubi} for a derivation of these equations):      
\begin{eqnarray}
\label{calbindin}
\nonumber
\frac{d[CaL_{0}]}{dt} & = & -4 k_{on1} [CaL_{0}]  [Ca^{2+}] + k_{off1} [CaL_{1}]\\
\nonumber
\frac{d[CaL_{1}]}{dt} & = & - k_{off1} [CaL_{1}]  - 3 k_{on1} [CaL_{1}]  [Ca^{2+}] + \\
\nonumber
&&
4  k_{on1} [CaL_{0}] [Ca^{2+}] + 2 k_{off1} [CaL_{2}]\\
\nonumber
\frac{d[CaL_{2}]}{dt} & = & - 2 k_{off1} [CaL_{2}]  -  2 k_{on1} [CaL_{2}]  [Ca^{2+}]+ \\
\nonumber
&&
3 k_{on1} [CaL_{1}] [Ca^{2+}] + 3k_{off1} [CaL_{3}] \\
\nonumber
\frac{d[CaL_{3}]}{dt} & = & -3 k_{off1} [CaL_{3}]  -  k_{on2} [CaL_{3}]  [Ca^{2+}] + \\
\nonumber
&&
2k_{on1} [CaL_{2}] [Ca^{2+}] + 4 k_{off2} [CaL_{4}] \\
\nonumber
\frac{d[CaL_{4}]}{dt} & = & -4 k_{off2} [CaL_{4}] + k_{on2} [CaL_{3}]  [Ca^{2+}]\\
\frac{ d[Ca^{2+}]}{dt}&=&-\sum_{n=1}^{4}n  \frac{ d[CaL_{n}]}{dt} +f([Ca^{2+}],t) 
\end{eqnarray}
where square brakets denote molar concentrations of each chemical species, CaL$_{i}$ with $i=0, 1, ..., 4$ denotes CaL with $i$ Ca$^{2+}$ bound ions, and $f([Ca^{2+}],t)$ is a function describing the time-dependent oscillations of Ca$^{2+}$ in the cell. We take the following values for the model parameters \cite{Nageri00}: $k_{on1}$=$1.3\cdot 10^{7}\, \mathrm{M}^{-1} \mathrm{s}^{-1}$, $k_{off1}$=$2.275\,  \mathrm{s}^{-1}$, $k_{on2}$=$7.7\cdot 10^{7}\, \mathrm{M}^{-1} \mathrm{s}^{-1}$, and $k_{off2}$=$39.501\, \mathrm{s}^{-1}$. The system of equations (\ref{calbindin}) has no analytical solutions and thus it must be solved with numerical methods. 

\begin{figure}
\includegraphics[width=4in]{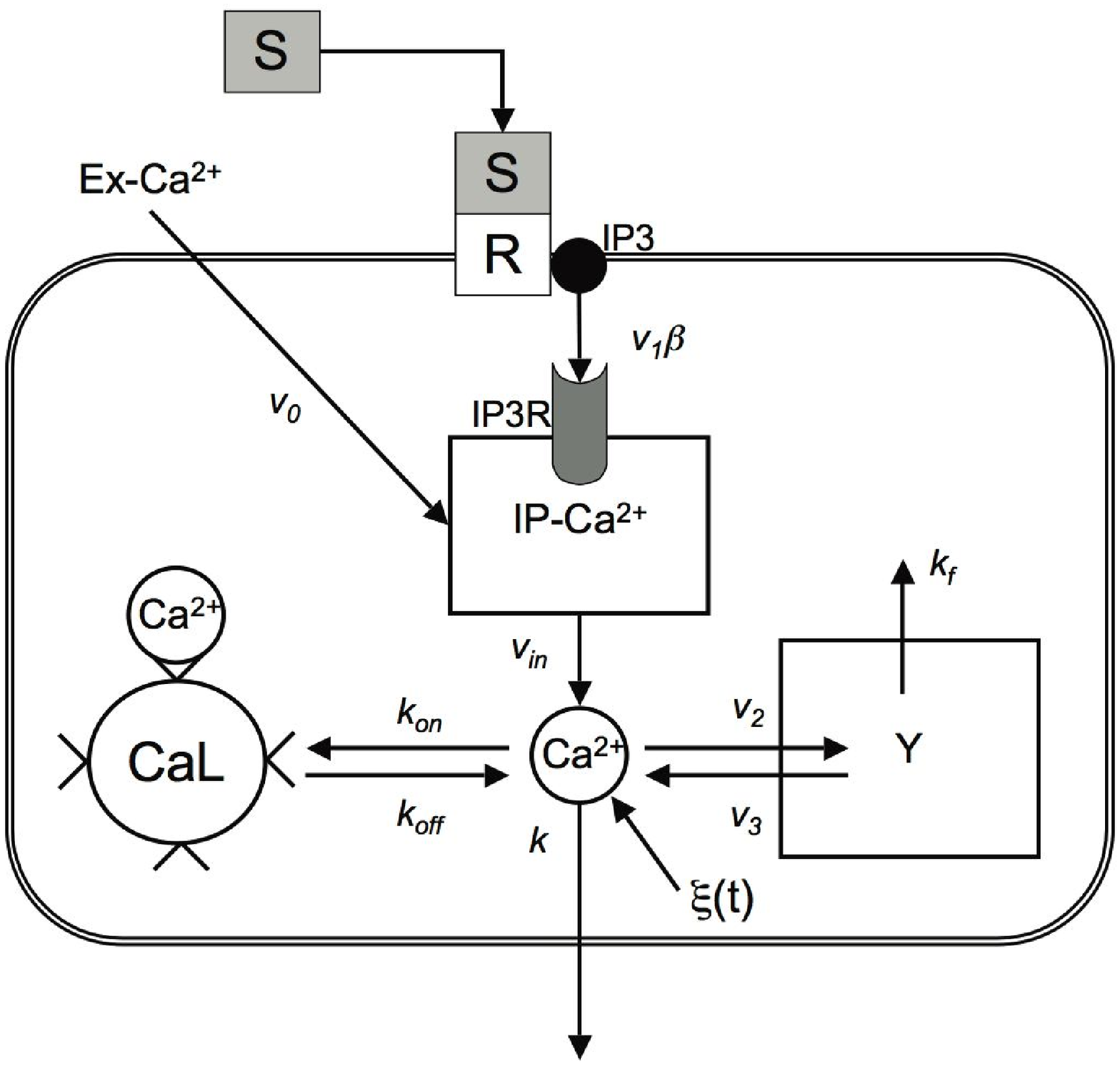}
\caption{\label{fig1}  Schematic representation of the CICR model. A signal S acts on receptor R and triggers the production of IP3 that stimulates the release of Ca$^{2+}$ ions from IP3-sensitive intracellular stores. IP3 regulates the constant flow of Ca$^{2+}$ into the cytosol ($v_1 \beta$).  Cytosolic Ca$^{2+}$ions are pumped into an IP3-insensitive store ($v_2$); Ca$^{2+}$  in this store (Y) is transported back to the cytosol in a process activated by Ca$^{2+}$ itself ($v_3)$. Rates $v_0$, $k$ and $k_f$ denote replenishment of the IP3-sensitive store with extracellular Ca$^{2+}$, the efflux of cytosolic Ca$^{2+}$ from the cell and passive leak of Ca$^{2+}$ from the store Y into the cytosol.  More details can be found in \cite{Falcke04}. This classical scheme is modified here to take into account the action of Ca$^{2+}$ protein buffers such as calbindin (CaL) that binds Ca$^{2+}$ ions on multiple sites. Binding follows the law of mass action and is described by rates $k_{on}$ and $k_{off}$. Finally, in our model cytosolic Ca$^{2+}$ fluctuates because of internal noise.}
\end{figure}

We still have to specify the function $f([Ca^{2+}],t)$; there are many different models of intracellular calcium oscillations, but this choice is not critical, and here we take the minimal and well-known model based on Ca$^{2+}$-induced Ca$^{2+}$ release (CICR) as the basic model of Ca$^{2+}$ oscillations \cite{Goldbeter90}, which is sketched in fig.\ref{fig1} (For recent reviews, see also \cite{Falcke04,sch} and references therein). 

The CICR model must be complemented by a stochastic term because 
the process of Ca$^{2+}$ wave generation is intrinsically noisy \cite{Yao95, Marchant01}.  Evidence of intracellular noise comes from direct inspection of sampled time-series in different cell types at the mesoscopic scale as well as from experimental work which shows that, in the microscopic domain, Ca$^{2+}$ waves originate from discrete random events, called puffs, occurring at specific sites in the cells and composed of a small number of Ca$^{2+}$ ions. Several puffs cooperate to the formation of a supercritical nucleus that initiatiates a Ca$^{2+}$ wave. The probabilistic character of nucleation introduces variability into the wave period, with a standard deviation which has been experimentally estimated to reach values up to 40\% \cite{Marchant01}. Thus, internal noise in Ca$^{2+}$ dynamics cannot be neglected, and we introduce a stochastic term (as in \cite{Li05, Hilborn05}  that leads to the following stochastic differential system for the driving function $f=f([Ca^{2+}],t)$:
\begin{eqnarray}
\label{calcium}
\nonumber
f = \frac{d[Ca^{2+}]}{dt} &=& v_0+v_1 \beta -v_2+v_3+k_f [Y] \\
\nonumber
&&
 -k [Ca^{2+}]+ \xi (t)\\
\frac{d[Y]}{dt} &=& v_2 - v_3 - k_f [Y]
\end{eqnarray}
where
\begin{equation}
\label{vel}
\nonumber
v_2 = V_{M2} \frac{[Ca^{2+}]^{n}}{(K_2^{n}+[Ca^{2+}]^{n})}
\end{equation}
and
\begin{equation}
\nonumber
v_3 = V_{M3} \frac{[Y]^{m}}{(K_R^{m}+[Y]^{m})} \frac{[Ca^{2+}]^{p}}{(K_A^{p}+[Ca^{2+}]^{p})}
\end{equation}
and $\xi (t)$ is Gaussian white noise with zero mean and variance $D$.

In these equations (see also fig.\ref{fig1}), 
$[Ca^{2+}]$ is the cytosolic Ca$^{2+}$ concentration, whereas $[Y]$ denotes Ca$^{2+}$ concentration in the IP3-insensitive intracellular store. $v_{0}$ = $10^{-6} \, \mathrm{M\, s}^{-1}$ is the input rate of Ca$^{2+}$ from the extracellular medium, $v_{1}$ = $7.3\cdot 10^{-6}\, \mathrm{M\, s}^{-1}$ is the parameter related to the IP3-modulated release of Ca$^{2+}$ from the IP3-sensitive store. $V_{M2}$ = $65\cdot 10^{-6}\, \mathrm{M\, s}^{-1}$  denotes the maximum rate of Ca$^{2+}$ pumping into the IP3-insensitive store, whereas  $V_{M3}$ = $500\cdot 10^{-6}\, \mathrm{M\, s}^{-1}$ is the maximum rate of release of Ca$^{2+}$ from that store into the cytosol in a process activated by cytosolic Ca$^{2+}$; $K_{2}$ = $10^{-6}\, \mathrm{M}$, $K_{R}$ = $2\cdot 10^{-6} \,\mathrm{M}$ and $K_{A}$ = $0.9\cdot 10^{-6} \,\mathrm{M}$ are threshold constants for pumping, release and activation, respectively; $k_{f}$ = $1 \,\mathrm{s}^{-1}$ is a rate constant that regulates the passive, linear leak of Ca$^{2+}$ from the IP3-insensitive store into the cytosol; $k$ = $10\, \mathrm{s}^{-1}$ regulates the assumed linear transport of cytosolic Ca$^{2+}$ into the extracellular medium. The exponents $n=m=2$ and $p=4$ denote the Hill coefficients characterizing these processes. 

The parameter $\beta$ regulates the saturation of the IP3 receptor and acts as the control parameter which sets the level of the stimulus and varies from $0$ to $1$.
We also introduce a parameter $\tau\ge1$ s that scales the rates $v_0, V_1, V_{M2}, V_{M3}, k_f$ and $k$. In this way we can tune the Ca$^{2+}$ period  of  the CICR oscillator to match oscillations observed in real experiments.

We use equations (\ref{calbindin}) and (\ref{calcium})  to investigate numerically the dynamic interplay between Ca$^{2+}$, Ca$^{2+}$-binding proteins and noise in the cell. We integrate the stochastic differential system with the Euler-Maruyama algorithm \cite{high}, and  for stability reasons we take a short integration time step $\Delta t = 0.001 s$.

\begin{figure}
\includegraphics[width=4in]{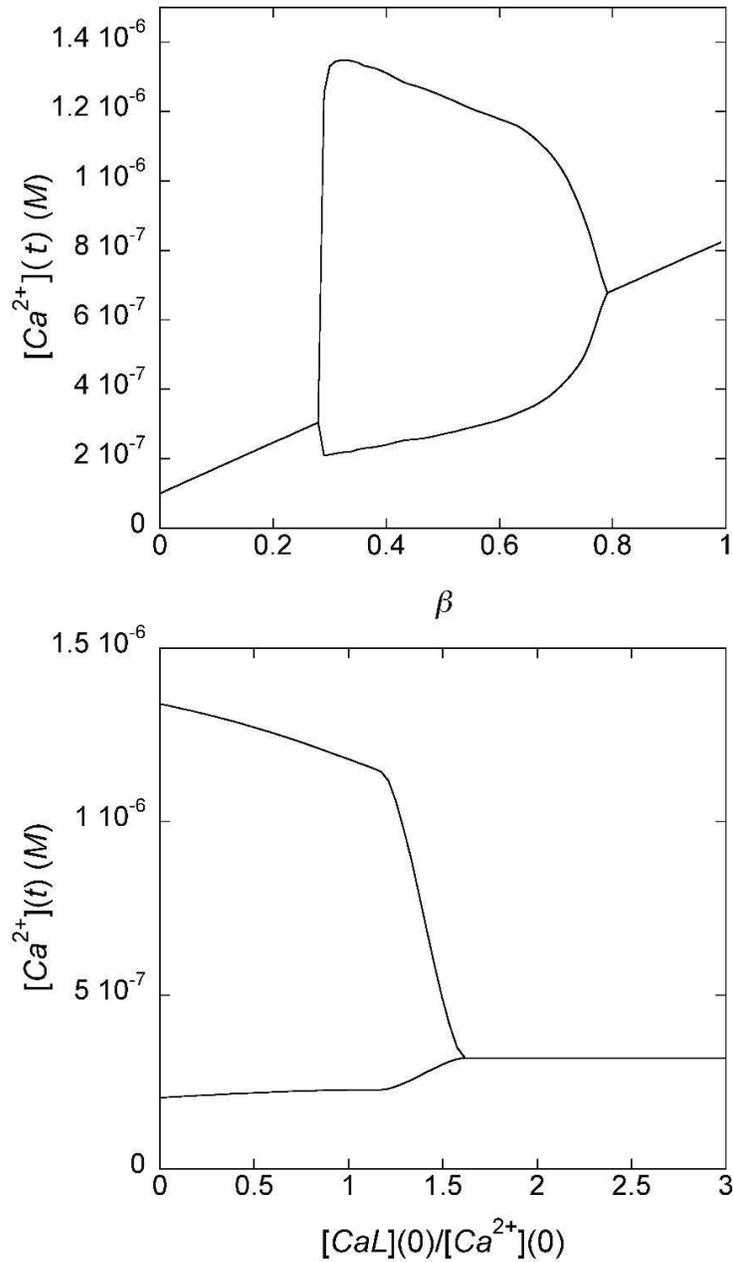}
\caption{\label{fig2}  Upper panel: Bifurcation diagram for the deterministic dynamics without CaL: in the absence of CaL the system displays two Hopf bifurcations as a function of the control parameter $\beta$. Lower panel: here $\beta$=0.3, which brings the system beyond the first Hopf bifurcation, into the oscillatory regime, however if CaL rises above a critical concentration ratio $[CaL](0)/[Ca](0)\approx1.6$, it blocks all oscillations. Thus, CaL sets a sharp threshold for Ca$^{2+}$ spiking.}
\end{figure}

\begin{figure}
\includegraphics[width=4in]{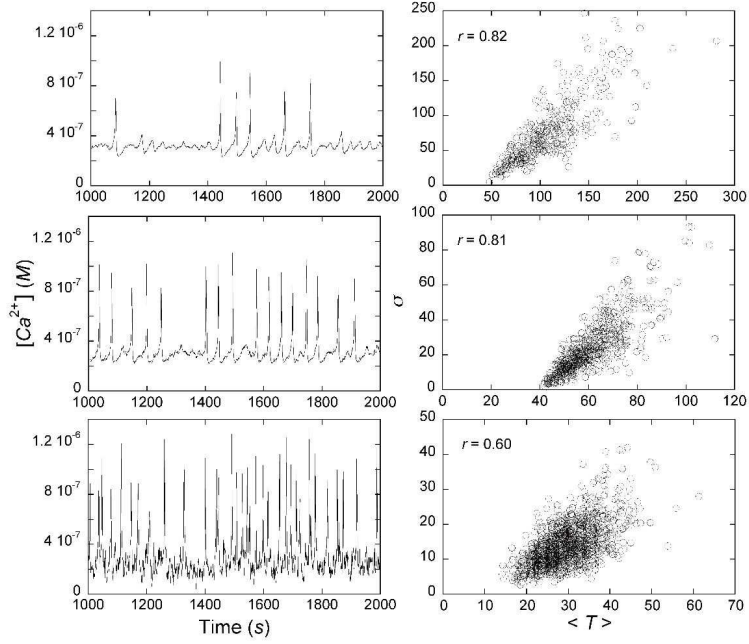}
\caption{\label{fig3}  Simulation outputs that show Ca$^{2+}$ spiking in presence of the Ca$^{2+}$-binding protein CaL and noise. In all panels the initial conditions are: $[Ca^{2+}](0)= 2\cdot 10^{-7}$ M, $[CaL](0)/[Ca^{2+}](0)=2$, $\beta=0.3$. All the other parameters are as described in the text. Top panels: $\log_{10}D=-16$ M$^2$. Middle panels: $\log_{10}D=-15.4$ M$^2$. Bottom panels:  $\log_{10}D=-13.6$ M$^2$. Time series are shown on the left whereas the plots on the right show the correlation between average interspike interval $\langle T\rangle$ and standard deviation $\sigma$. These simulation results are strikingly similar to those found in actual experiments \cite{Skupin08}}  
\end{figure}

Before studying the influence of internal noise, we investigate the deterministic system's dynamics both in the absence and in the presence of CaL. Simulations show that in the absence of CaL, the system undergoes two Hopf bifurcations at $\beta\approx0.29$ and $\beta\approx0.77$. Increasing concentrations of CaL damp the Ca$^{2+}$ oscillations, until they abruptly switch off at the critical concentration $[CaL](0)/[Ca^{2+}](0)\approx1.6$ (with $\beta = 0.3$, see fig.\ref{fig2}). 
Beyond this critical CaL concentration no Ca$^{2+}$ oscillations are observed in the deterministic case. However, fluctuations of Ca$^{2+}$ concentration due to internal noise allow the system to cross the threshold again and oscillate even in the presence of supercritical CaL concentrations (fig.\ref{fig3}). One feature of the simulation outputs in fig.\ref{fig3} is that they are strikingly similar to time series sampled in real cells \cite{Skupin08}. In these experiments, Ca$^{2+}$ spikes have been observed to occur randomly, and the average interspike interval $\langle T\rangle$ measured independently for hundreds of cells is correlated to the standard deviation $\sigma$ in a rather broad range of $\langle T\rangle$ values \cite{Skupin08}. We find the same result in our simulations (fig.\ref{fig3}).

\begin{figure}
\includegraphics[width=4in]{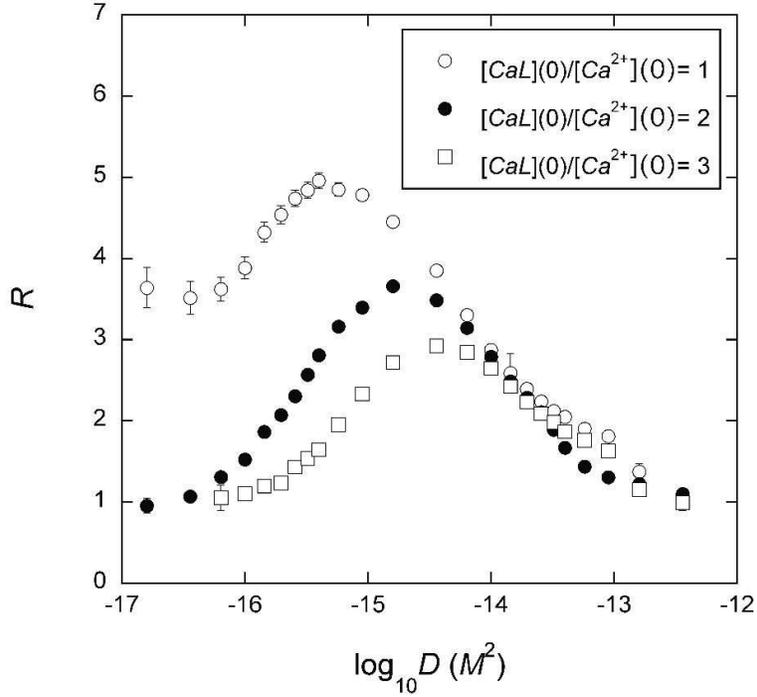}
\caption{\label{fig4}  Plot of the regularity parameter $R$ vs. noise variance for the indicated $[CaL](0)/[Ca^{2+}](0)$ ratios. Since both Ca$^{2+}$ and CaL concentrations vary greatly among different cell types, we use the ratios $[CaL]/[Ca^{2+}]$ rather than concentrations. The presence of CaL deeply influences the nature of the calcium oscillator and leads to a stabilization of interspike intervals: this regularization of the Ca$^{2+}$ oscillations depends on noise variance and reaches a maximum for a nonvanishing value of the noise variance.}
\end{figure}

Thus, both experiments and the numerical results shown in fig.\ref{fig3}, suggest that noise-induced Ca$^{2+}$ spiking and spiking periodicity depend on noise variance, and moreover the observed correlation between noise variance and oscillation frequency also suggest that what we are witnessing here is a form of stochastic coherence \cite{Hilborn05}. As in \cite{Hilborn05} we study this aspect using the regularity parameter $R$, which is the ratio between the average and standard deviation of the interspike interval (the ``period'' of the oscillations):
\begin{equation}
\label{regularity}
R = \frac{\langle T\rangle}{\sigma}
\end{equation} 
As already noted in \cite{Hilborn05}, the reciprocal of this quantity is just the ``coefficient of variation'' often used in neuroscience as an estimator of the regularity  interspike intervals.
In our simulations (see fig.\ref{fig4}) we find that regularity peaks at a given noise variance, and that the position of the peak also depends on the CaL concentration. Thus the whole system displays stochastic coherence, and proteins that bind Ca$^{2+}$ ions on multiple sites can tune Ca$^{2+}$ spiking in cells and may ultimately regulate cell communication. Remarkably, experiments with engineered knock-out mice for CaL expression show severe impairment of motor coordination suggesting that CaL has an important role in cerebellar functions and intercellular communication \cite{Airaksinen97}. We conclude that proteins, such as CaL, may not simply act as passive  Ca$^{2+}$ buffers in cells, but be prime actors in the complex play of cellular communication.

\bibliography{apssamp}

\pagebreak


\newpage 
\bibliography{apssamp}

\end{document}